\documentclass[12pt]{article}
\usepackage{amsmath,amssymb}
\usepackage{graphicx,color}
\usepackage{cite}
\usepackage{bm}
\usepackage{dcolumn}
\newcommand{\be}{\begin{equation}}
\newcommand{\bea}{\begin{eqnarray}}
\newcommand{\eea}{\end{eqnarray}}
\newcommand{\ba}{\begin{array}}
\newcommand{\ea}{\end{array}}
\newcommand{\ee}{\end{equation}}

\expandafter\ifx\csname mathbbm\endcsname\relax

\else

\fi \textheight 22cm \textwidth 15cm \topmargin 1mm
\begin{document}

\begin{titlepage}
\hfill
\vbox{
    \halign{#\hfil         \cr
           IPM/P-2005/086 \cr
                      } 
      }  
\vspace*{20mm}
\begin{center}
{\Large {\bf On 1/2 BPS Solutions in M-theory}\\ }

\vspace*{15mm}
\vspace*{1mm}
{Mohsen Alishahiha \footnote{alishah@theory.ipm.ac.ir}
and Hossein Yavartanoo }
\footnote{yavar@ipm.ir} \\
\vspace*{1cm}

{\it  Institute for Studies in Theoretical Physics
and Mathematics (IPM)\\
P.O. Box 19395-5531, Tehran, Iran \\ \vspace{3mm}}

\vspace*{1cm}
\end{center}

\begin{abstract}
We study singular 1/2 BPS solutions in M-theory using 11-dimensional superstar solutions.
The superstar solutions and their corresponding  plane wave limits could give an insight
how one may deform the boundary conditions to get singular, but still physically acceptable,
solutions. Starting from M-theory solutions with an isometry, we will also study 10-dimensional 
solutions coming from these M-theory solutions
compactified on a circle.

\end{abstract}
\end{titlepage}
\section{Introduction}

The AdS/CFT correspondence \cite{{Maldacena:1997re},{Gubser:1998bc},{Witten:1998qj}} 
is an explicit example of holographic principle which might increase our knowledge about
what a fundamental theory of quantum gravity could be. In this framework a quantum mechanical 
system which includes gravity can be described by a lower dimensional quantum mechanical system 
without gravity. Having had gravity on one side of the duality one may then wonder if we can learn 
about quantum gravity by studying a quantum field theory which by now we have more control on it.

We note, however, that this correspondence correspondence is a strong/weak duality in the sense 
that when the gauge theory is perturbative the string theory sigma model is
strongly coupled and vice-versa. Hence usually one can only perform perturbative computations 
in one side of the duality which makes the comparison of two sides difficult. Nevertheless 
it has recently been shown that there is a certain large quantum numbers limit, the BMN sector 
\cite{BMN} and the ``semi-classical'' strings \cite{Gubser:2002tv,Frolov:2002av}, in
which the gauge theory and the string theory sides are both
perturbatively accessible and therefore might be used to study this duality better.

To be specific let us consider $AdS_5\times S^5$ case in the global coordinates whose gauge theory
dual is an ${\cal N}=4$ SYM on $R^1\times S^3$. Then one may consider 1/2 BPS states of the gauge 
theory which could be given by operators carrying R-charge $J$ with dimension $\Delta$ such that
$\Delta=J$. From gravity dual point of view these BPS states correspond to the gravity modes for small 
energy; $J\ll N$, while for the energy of order of $N$ they correspond to the brane configurations which can in
principle cause geometric changes in $AdS_5\times S^5$ due to back reaction.

In order to find the geometric counterparts of this class of
operators, the authors of \cite{Lin:2004nb} have established the
general setting for obtaining the corresponding $1/2$ BPS
geometries. This is done by looking for those solutions of type
IIB supergravity equations which have $SO(4)\times SO(4)\times R$
isometry and which solve the killing spinor equations. Doing so,
one picks from all the excitations in $AdS_5\times S^5$ in the
global coordinates, those which constitute its $1/2$ BPS sector.
It turns out that these symmetry requirements, plus regularity,
are very restrictive such that the whole solution is determined by
a single function which should satisfy a linear differential
equation subject to certain boundary conditions on a 2-plane. 

This construction has also  been generalized for M-theory in \cite{Lin:2004nb} where the authors have 
considered 1/2 BPS geometries of M-theory which have $SO(3)\times SO(6)\times R$ isometry. In this case
the solution can be fixed by solving a three dimensional Toda equation. To get a non-singular
solution one needs to put a specific boundary condition. In the next section we will review
this construction.  

In contrast to the 1/2 BPS geometries in type IIB string theory which can be obtained by solving 
the Laplace equation which is a linear differential equation, in M-theory we have to deal with a non-linear
operator which is the three dimensional Toda equation. Thus it is very difficult to study
1/2 BPS states in M-theory, as we can not find a new solution by superposition of other known
solutions such as $AdS_{7,4}\times S^{4,7}$. Therefore it would be interesting to find any
possible solutions of this Toda equation. Different aspects of 1/2 BPS in M-theory have also
been studied in \cite{{Suryanarayana:2004ig},{Bak:2005ef},{MacConamhna:2005vt},{Spalinski:2005ha},{Lin:2005nh}, 
{Milanesi:2005tp},{Ganjali:2005cr}}.  It is the aim of this paper to further study 1/2 BPS states in M-theory. We shall
consider both singular and smooth solutions.

The organization of the paper is as follows. In section 2 we shall review the 1/2 BPS geometries
in M-theory. We will also study well-known solutions of M-theory using different coordinates systems
from that we could understand different aspects of M-theory 1/2 BPS solutions.
In section 3 we will consider superstar solutions in M-theory in the limit where we get 
singular 1/2 BPS solutions. We will also study their Penrose limit using LLM parametrization.
In section 4 using what we have learned from superstars solutions we shall study different
possible boundary condition which could lead to either singular and smooth solutions.
In section 5 we will study some 10-dimensional solution by making used by compactification of those
M-theory solutions which have an isometry. The last section is devoted to conclusion.  

\section{1/2 BPS geometries in M-theory}
In this section we fix our notation and review 1/2 BPS geometries in M-theory 
constructed in \cite{Lin:2004nb} where 1/2 BPS geometries with $SO(3)\times SO(6)\times R$
have been classified. These geometries can be thought of as  1/2 BPS geometries in $Ads_7\times S^4$.
These could be associated to the chiral primaries of the $(2,0)$ theory.

Using the Killing spinor equation the authors of \cite{Lin:2004nb} have been able to find
M-theory supergravity solutions which preserve 16 supercharges and have $SO(3)\times SO(6)$
isometry 
\bea 
\label{4d_general} 
ds_{11}^2&=&-{ 4e^{2\lambda} ( 1
+ y^2e^{-6\lambda} ) } (dt+V_idx^i)^2 + \frac{ e^{-4\lambda} }{1+y^2e^{-6\lambda}}
[ dy^2 + e^{ D} (dx_1^2 + dx_2^2) ]\cr
&&+4e^{2\lambda} d\Omega_5^2+y^2e^{-4\lambda}d{\tilde\Omega}_2^2,\cr
 e^{-6\lambda} &=&\frac{\partial_y  D }{ y(1 -  y \partial_y D) },\;\;\;\;
 \;\;\;\;\;
V_i = \frac{1}{2} \epsilon_{ij} \partial_j. 
\eea
where $i,j=1,2$. There is also a four form which we have not written it (For
detail see \cite{Lin:2004nb}). The solution is completely determined by a function
$D(y,x_1,x_2)$ which satisfies the three dimensional Toda  equation
\be \label{toda}
(\partial_1^2 + \partial_2^2 )D + \partial_y^2 e^{D} =0.
\ee
The solution is fixed as soon as the boundary condition for this Toda equation is given. 
One may 
further put a boundary condition such that the obtained solution is non-singular.
To do that we note that the radii of the five and two spheres are related to
$y$ coordinate, $y=R_2R_5^2/4$. This means that at $y=0$ either the two sphere or 
five sphere shrinks to zero. Therefore the boundary condition must be such that
the geometry become non-singular when either of the spheres shrink.
One can see that the geometry is non-singular if we impose the following
boundary condition \cite{Lin:2004nb}
\bea \label{boundary1}
{\rm At}&&\;\;\;y=0,\;\;\;\;\;   \partial_y D=0,\;\;\;\;\;D={\rm finite},\;\;\;\;\;\;\;\;\;S^2 
\;\;{\rm shrinks},\cr
{\rm For}&&\;\;\;y\rightarrow 0,\;\;\;\;\;e^D \sim \;y,\hspace{3.7cm} S^5 \;\;{\rm shrinks}.
\eea
This can separate the 12 plane into droplets where we have one or the other boundary conditions.

\begin{figure}
 \begin{center}
  \includegraphics[scale=.6]{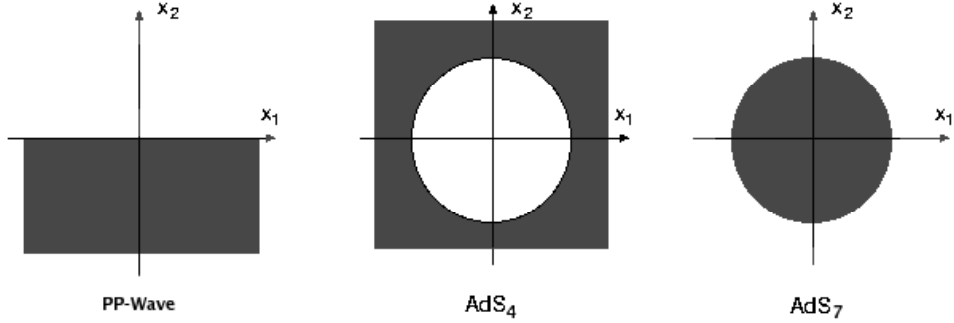} 
 \end{center}
\caption{pp-wave and AdS$_{7,4}$ boundary conditions}
\label{pp-wave:fig} 
\end{figure}

Let us write down some specific examples. For simplicity we assume that the solution has isometry in
$x_1$ direction and therefore we will need to solve a two dimensional Toda equation. 
The simplest example is the plane wave solution in  which the boundary condition is given by 
(see figure \ref{pp-wave:fig})
\bea
\partial_y D &=&0 , \;\; D={\rm finite}  \;\;\; {\rm at}\;\;\;y=0,\;\;\;\;\;\;{\rm in}\;\;\;x_2 > 0 ,\cr
e^D &\sim&  y \;\;\;\;\;\;\;\;\;\;\;\;\;\;\;\;\;\;\;\;\;
{\rm for}\;\;\;y=0,\;\;\;\;\;\;{\rm in}\;\;\; x_2 < 0.
\eea
Since the solution has an isometry in $x_1$ direction it can be parametrized by two
parameters $r_5$ and $r_2$ which are
the radial coordinates in the first six transverse dimensions and
the last three transverse dimensions, respectively. In this parametrization the plane wave solution
is given by 
\be
\label{mthpp}
y = \frac{ 1}{4} r_5^2 r_2,\;\;\;\;\; 
x_2 =\frac{ r_5^2}{4 } - \frac{r_2^2}{ 2},\;\;\;\;\;\;
e^{D}=\frac{r_5^2}{4}.
\ee 

As another example let us consider  $AdS_7 \times S^4$ solution 
in which going to polar coordinates $(x,\phi)$ the solution has an isometry in $\phi$ direction and 
therefore can be expressed by two parameters as follows \cite{Lin:2004nb}
\be
\label{ads7}
e^D = \frac{r^2L^{-6}}{4+r^2}, \;\;\;\;\; x=(1+\frac{r^2}{4})\cos \theta, \;\;\;\;\; 4y=L^{-3}
r^2\sin \theta,
\ee
where $\theta$ is an angular direction on $S^4$ and $r$ is the radial coordinate in AdS$_7$ and
$L$ is the radius of $S^4$. In terms of $x$ coordinate the corresponding boundary 
condition may be written as 
\bea
\partial_y D &=&0 , \;\; D={\rm finite}  \;\;\; {\rm at}\;\;\;y=0,\;\;\;\;\;\;{\rm in}\;\;\;x > 1 ,\cr
e^D &\sim&  y \;\;\;\;\;\;\;\;\;\;\;\;\;\;\;\;\;\;\;\;\;
{\rm for}\;\;\;y=0,\;\;\;\;\;\;{\rm in}\;\;\; x <1,
\label{ads7b}
\eea
which could be thought of as a black disk in 12 plane (see figure \ref{pp-wave:fig}).
 Similarly we can describe the solution for $AdS_4\times S^7$ as follows
\bea
\label{ads4}
e^{D}=4L^{-6}\sqrt{1+\frac{r^2}{4}}\sin^2\theta,\qquad
x=\left(1+\frac{r^2}{4}\right)^{1/4}\cos\theta,\qquad
2y=L^{-3}r\sin^2\theta, 
\eea 
where $L$ is the radius of AdS$_4$. The boundary condition is given by
\bea
\partial_y D &=&0 , \;\; D={\rm finite}  \;\;\; {\rm at}\;\;\;y=0,\;\;\;\;\;\;{\rm in}\;\;\;x<1 ,\cr
e^D &\sim&  y \;\;\;\;\;\;\;\;\;\;\;\;\;\;\;\;\;\;\;\;\;
{\rm for}\;\;\;y=0,\;\;\;\;\;\;{\rm in}\;\;\; x>1,
\label{ads4b}
\eea
which again represents a circular droplet in 12 plane, though in this case it is a white
disk in a black plane (see figure \ref{pp-wave:fig}).

We note that in both $AdS \times S$ cases we have circular
droplets and the solutions are exchanged by exchanging the black white regions. In fact the situation is 
very similar to AdS$_5$ solution in type IIB where the corresponding boundary condition is given by 
droplet which is a black disk in 12 plane. In this case the theory has two $Z_2$ symmetries, both of which
exchange black and white regions \cite{Alishahiha:2005fc} (see also \cite{Lin:2005nh}). 
One of them which can be interpreted as the particle-hole
symmetry is the symmetry of the whole supergravity solution though the second one is just
the symmetry of the metric and changes the sign of five from flux.

On the other hand in M-theory we could still exchange the black and  white regions, though we would not 
expect that this leads to any symmetry. In fact under this action, as shown in figure \ref{pp-wave:fig},
AdS$_7$ maps to AdS$_4$. Actually, this can be interpreted as electric-magnetic duality in which
M2-brane maps to M5-brane. One note, however, that the plane wave solution is invariant under this
map which is acts as $x_2\rightarrow -x_2$. This if of course expected because both AdS$_7$ and
AdS$_4$ lead to the same plane wave solution by taking Penrose limit \cite{Berenstein:2002jq}.

To further compare these two solutions, and in fact other possible solutions which 
we will study in the next section, it is useful to define new coordinates 
$x_2$ as $x=e^{x_2}$ and also consider a
new parameter $\xi$ such that $\frac{r}{ 2}=\sinh\xi$. In this notation the $AdS_7$ reads
\be
y=L^{-3}\sinh^2\xi\;\sin\theta,\;\;\;\;x_2=2\ln\cosh\xi+\ln\cos\theta,\;\;\;\;
e^{\tilde{D}}=\frac{L^{-6}}{4}\sinh^22\xi\;\cos^2\theta.
\label{lll}
\ee
Here we have used the fact that the solution of the Toda equation is invariant under a conformal
transformation which in our case we have $x=e^{x_2},\;\;D=\tilde{D}-2x_2$. Similarly for $AdS_4$ 
case one finds\footnote{In order to make the equation similar to $AdS_7$ case we have also
rescale $x_2\rightarrow 2x_2$ and $e^{\tilde{D}}\rightarrow e^{\tilde{D}}/4$.}
\bea
y=L^{-3}\sinh\xi\;\sin^2\theta,\;\;\;\;x_2=\ln\cosh\xi+2\ln\cos\theta,\;\;\;\;
e^{\tilde{D}}=\frac{L^{-6}}{4}\sin^22\theta\;\cosh^2\xi.
\eea
These two solutions are exchanged under $\xi\leftrightarrow i\theta$ which is expected 
as $\xi $ and $\theta$ represent 
$AdS_{7,4}$ in the global coordinates. Writing the solution in this coordinate is useful specially if one wants
to study plane wave limit of the solutions. 

We note also that writing $AdS_{7,4}$ in the $(x_2,y,\tilde{D})$ coordinates, the boundary
condition we get is similar to that we have in plane wave solution ( figure 1). This is of course
expected as the conformal map $x=e^{x_2}$ maps a disk to a plane. This might look puzzling taking 
into account that both plane wave and $AdS_{7,4}$ seem to have the same boundary condition at $y=0$.
We note however that although the map $x=e^{x_2}$ maps a disk to plane, since the disk includes 
$x=0$ point, which will be mapped to $x_2=-\infty$ and therefore the boundary condition we get at the
end of the day is defined on a plane which includes $x_2=-\infty$ as well. This is the fact
that distinguishes between $AdS$ and plane wave solutions.\footnote{We would like thank 
M.M. Sheikh-Jabbari for a discussion
on this point.} Therefore we conclude that the topology of the 12 plane is indeed important.



Let us now study the Penrose limit of the $AdS_{7,4}\times S^{4,7}$ using the 
LLM parametrization.\footnote{The Penrose limit of these background which leads to the same
plane wave in M-theory has been studied in \cite{{Berenstein:2002jq},{Blau:2002dy}}.} 
Using the LLM parametrization is also useful to study the next order corrections to the plane wave background.
These corrections for type IIB string they in the LLM language has been studied in \cite{Takayama:2005bc}.
The Penrose limit of different backgrounds in LLM context has also been studied in 
\cite{Ebrahim:2005uz}.

\begin{figure}
 \begin{center}
  \includegraphics[scale=.25]{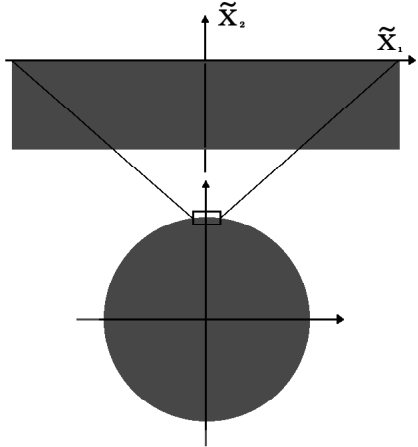} 
 \end{center}
\caption{Penrose Limit}
\label{pp-wave2:fig} 
\end{figure}

The procedure of taking the Penrose limit of a given solution in LLM context is as follows. In fact
from LLM point of view this can be done by focusing on a small region around the edge of
a droplet in the $(x_1,x_2)$ plane and then blowing up this region.
The boundary condition we get  corresponds to the plane-wave
solution (see figure \ref{pp-wave2:fig}).
We note, however, that, as we saw in the previous section, having a conformal symmetry the shape of the
droplet is not really important. 
In particular for the cases we have considered, all of them
have the same shape, though in the $AdS_{4,7}$ one needs to add to the plane the $x_2=-\infty$ point.
Therefore in order to find the plane wave solution in this parameterization, one need to look at
a region which are far from $x_2=-\infty$ point and therefore effectively we would see
the plane wave boundary condition and thereby we will get plane wave solution.

Let us first study this procedure for the $AdS_7\times S^4$ solution. To imply the above 
procedure we will consider the limit of $\xi\rightarrow 0,\;\theta\rightarrow 0$. To be
precise we will consider a limit in which $L\rightarrow 0$ while keeping the following quantities fixed
\be
r_5=\frac{\xi}{L^3},\;\;\;\;\;r_2=\frac{\theta}{L^3},\;\;\;\;\;{\tilde y}=\frac{y}{L^6},\;\;\;\;\;{\tilde x}_2=
\frac{x_2}{L^6}.
\label{li7}
\ee
In this limit the solution (ref{lll}) reads
\be
\tilde{y}=r_5^2r_2,\;\;\;\;\;\;\;\;\;\tilde{x}_2=r_5^2-\frac{1}{2}r_2^2,\;\;\;\;\;\;\;\;e^{\tilde{D}}=r_5^2,
\ee
which is the plane wave solution (\ref{mthpp}) up to a rescaling. The procedure goes the same for
$AdS_4\times S^7$ where upon defining $r_2=\frac{\xi}{L^3},r_5=\frac{\theta}{L^3}$ we get
\be
\tilde{y}=r_5^2r_2,\;\;\;\;\;\;\;\;\;\tilde{x}_2=\frac{1}{2}r_2^2-r_5^2,\;\;\;\;\;\;\;\;e^{\tilde{D}}=r_5^2,
\ee
which is equivalent to the above plane wave solution taking into account the $x_2\rightarrow -x_2$
symmetry.
It is of course well know that both $AdS_4\times S^7$ and $AdS_7\times S^4$ lead to the same 
plane wave solution under performing the Penrose limit.

\section{Giant graviton deformation: Superstar}

\subsection{Superstar solution}
In the previous section we have studied the simplest examples of 1/2 BPS in M-theory, including
plane wave, $AdS_7\times S^4$ and $AdS_4\times S^7$. From LLM point of view these solutions
can completely be given as soon as the boundary condition on 12 plane is specified. For the solutions
we have considered these boundary conditions are depicted in figure \ref{pp-wave:fig}.
We have also seen that there is a ``$Z_2$'' transformation under which the boundary condition
for $AdS_7$ maps to that for $AdS_4$ while the plane wave solution maps to itself. It is also expected
because both $AdS_7$ and $AdS_4$ lead to the same 11-dimensional plane wave. 

One may deform these solutions by adding spherical M5/M2 branes wrapped on $S^{5,2}$ spheres. This corresponds
to deform the theory by giant gravitons. Since the ``$Z_2$'' map exchanges the role of $S^5$ and $S^2$, this
corresponds to exchanging M5 and M2 branes or giant gravitons with dual giants. Following $AdS_5\times S^5$
case in type IIB string theory, adding giant gravitons might lead to either singular or smooth solution.
Of course since in the M-theory case we will have to deal
with a three dimensional Toda equation and because it is not a linear equation, finding such a solution
is very difficult.  

Fortunately the singular deformation has already been studied in the literature and known as
superstar \cite{{Leblond:2001gn},{Balasubramanian:2001dx}}. We can use this solution to 
study some feature of giant gravitons in M-theory using LLM construction.

Let start from supergravity solution representing giant gravitons on $AdS_7\times S^4$ background.
The corresponding solution is \cite{{Cvetic:1999xp},{Leblond:2001gn},{Balasubramanian:2001dx}}
\bea
ds_{11}^2& = & \Delta^{1/3}\Bigl(-(H_1 \, H_2)^{-1}f\,
dt^2+(f^{-1}dr^2+r^2 d\Omega_5^2)\Bigr)\\ && \,\,\,\,\,\,\,
+\frac{1}{4} \Delta^{-2/3}\Bigl[L_7^2 d\mu_0^2+\sum_{i=1}^2
H_i\Bigl(L_7^2 \, d\mu_i^2+\mu_i^2(L_7 \, d\phi_i+
\frac{\tilde{q}_i}{r^4+q_i}dt)^2\Bigr)\Bigr] \, ,
\eea
where $f=1-\frac{\mu}{r^4}+\frac{r^2}{L_7^2}(H_1H_2),\;\;H_i=1+\frac{q_i}{r^4}\;\;
{\tilde q}_i=\sqrt{q_i(\mu-q_i)}$  and
\be
\Delta \equiv (H_1 H_2)\Bigl(\mu_0^2+\sum_{i=1}^2 \frac{\mu_i^2}{H_i}\Bigr),  ~~~~
\mu_0 \equiv \sin\theta_1 \sin \theta_2, ~~~~ \mu_{1} \equiv \cos\theta_{1},~~~~ \mu_{2} 
\equiv \sin\theta_{1}\,
\cos\theta_{2}\,,
\ee
with $L_7^2=4 l_p^2 (\pi N)^{2/3}$.  This solution is
asymptotically $AdS_7 \times S^4$ with radius $L_7$. $N$ counts the number of M5-branes 
whose near-horizon limit is $AdS_7 \times S^4$. There is also a nonzero six form (dual to a three form)
which we have not written it. Of course we are interested in
the case where the solution is 1/2 BPS which can be obtained by setting $q_2=\mu=0$ and $q_1=Q$.
In this case, setting $\psi=\phi+t/L_7$, the solution reads
\bea
\label{metric0}
ds^2&=&-\Delta^{1/3}H_1^{-1}fdt^2+\frac{\Delta^{-2/3}H_1}{4}\cos^2\theta_1(L_7d\psi-H_1^{-1}dt)^2
+\Delta^{1/3}f^{-1}dr^2\cr && +\frac{L_7^2}{4}\Delta^{1/3}d\theta_1^2
+\Delta^{1/3}r^2d\Omega_5^2+\frac{L_7^2}{4}\Delta^{-2/3}\sin^2\theta_1 d\Omega_2^2,
\eea
which can be recast to the following form of 1/2 BPS solution of M-theory 
\bea
ds^2&=& -4e^{2\lambda}(1+y^2e^{-6\lambda}) (d\hat{t} +V_{\psi}d\psi)^2  +\frac{e^{-4\lambda}}{1+y^2e^{-6\lambda}}\left(dy^2+e^D(dx^2+x^2d\psi^2) \right)
\cr\cr && + 4e^{2\lambda} d\Omega_5^2 + y^2e^{-4\lambda} d\Omega_2^2, 
\label{standard}
\eea
with
\be
e^{2\lambda} = \frac{r^2}{4} \Delta^{1/3},\;\;\;\;\;\;\;V_{\psi}= \frac{-2L_7^2\cos^2\theta_1}
{r^2 \Delta + 4L_7^2 \sin^2\theta_1},
\ee
where
\be
x= g(r) \cos \theta_1 \;, \;\;\;\;\;\;\;\;\;\;\;  y= \frac{L_7}{8}{r}^2 \sin\theta_1 \;, \;\;\;\;\;\;\;\;\;\;\;
e^D= \frac{L_7^4r^2 f}{64 g^2}. 
\label{super7}
\ee
Note that the solution is given by two functions $f$ and $g$ such 
that $\frac{g'}{g} = \frac{2{r}}{L_7^2f}$, where 
in our notation $f=1+{r}^2/L_7^2+Q/{r}^2L_7^2$.

To summarize we saw that the superstar solution in M-theory is given by the metric 
(\ref{4d_general}) with the parametrization of (\ref{super7}).
From LLM language this solution can be obtained by
solving the three dimensional Toda equation with a boundary condition which should not be in the
form of 
(\ref{boundary1}) and therefore the obtained solution is singular. For this particular example we find the
following boundary condition.
\bea
\theta=0&:&\;\;\;\;x>x_0,\;\;\;\;\;\partial_yD=0,\;\;D={\rm finite},\cr &&\cr
{r}=0&:&\;\;\;\;x<x_0,\;\;\;\;e^D\sim c_0 (L_7^2+\frac{L_7^4}{Q}y),
\eea
where $c_0=(-4Q/(L_7^2+\sqrt{L_7^4-4Q})^2)^{L_7^2/\sqrt{L_7^4-4Q}}$ and $x_0=\sqrt{-Q/c_0}$. 
We note that this boundary condition, in the case of $r=0$, differs
from that in (\ref{boundary1}) and therefore the solution is singular.

Let us now study a giant graviton deformation of $AdS_4\times S^7$. The corresponding supergravity
solution is given by \cite{{Cvetic:1999xp},{Leblond:2001gn},{Balasubramanian:2001dx}}
\bea
ds_{11}^2& = &
\Delta^{2/3}\Bigl(-(H_1H_2H_3H_4)^{-1}f\, dt^2+(f^{-1}dr^2+r^2
d\Omega_2^2)\Bigr)\\ && \,\,\,\,\,\,\,
+4\Delta^{-1/3}\sum_i H_i\Bigl(L_4^2 \,d\mu_i^2+\mu_i^2 \, (L_4 \, d\phi_i+
\frac{\tilde{q}_i}{r+q_i}dt)^2\Bigr) \,,
\eea
where $f=1-\frac{\mu}{r}+\frac{r^2}{L_4^2}(H_1H_2H_3H_4),\;\;H_i=1+\frac{q_i}{r},\;\;
\Delta=(H_1H_2H_3H_4)\sum_{i=1}^4\frac{\mu_i^2}{ H_i}$ and 
\be 
\mu_1=\cos\theta_1, ~
\mu_2=\sin\theta_1\cos\theta_2, ~
\mu_3=\sin\theta_1\sin\theta_2\cos\theta_3, ~ {\rm and} ~
\mu_4=\sin\theta_1\sin\theta_2\sin\theta_3. 
\ee
There is also a nonzero three form. The solution is asymptotically ${\rm AdS}_{4} \times
{\rm S}^{7}$ where the AdS and sphere length scales are $L_4$ and
$2L_4$ respectively.  We are particularly interested in the case where only one charge is non-zero, 
and the solution is  in the extremal limit. This leads to a 1/2 BPS solution in 11-dimensions. 
In this case, setting $q_2=\mu=0,\;q_1=Q$ and $\psi=\phi_1+t/L_4$, the metric reads  
\bea
\label{metric1}
ds_{11}^2& = &
-\Delta^{2/3}H_1^{-1}f dt^2+4\Delta^{-1/3} \cos^2\theta_1 H_1 (L_4 d\psi-H^{-1}dt)^2
+\Delta^{2/3}f^{-1}dr^2\cr &&\cr
&&+4\Delta^{2/3} L_4^2d\theta_1^2+\Delta^{2/3}r^2
d\Omega_2^2+ 4\Delta^{-1/3} L_4^2 \sin^2\theta_1 d\Omega_5^2,
\eea
which can be recast to the standard form as (\ref{standard}) with
\be
e^{2\lambda}=L_4^2\Delta^{-1/3}\sin^2\theta_1,\;\;\;\;\;\;\;\;\;\;\;\;\;
V_{\psi}=  \frac{4L_4^2\cos^2\theta_1}
{{r}^2 \Delta + 4L_4^2 \sin^2 \theta_1 \;},
\ee
where
\be
x= g(r) \cos \theta_1 \;, \;\;\;\;\;\;\;\;\;\;\;  y= L_4^{2}r \sin^2\theta_1 \;,\;\;\;\;\;\;\;\;\; 
e^D =  \frac{ 4 L_4^6 f  \sin^2 \theta_1}{g ^2}.
\label{super4}
\ee
The solution is given in terms of two functions $f$ and $g$ such that $\frac{g'}{g}=\frac{r}{2L_4^2f}$ with
$f=1+r^2/L_4^2+Qr/L_4^2$.

To summarize we note that the giant graviton deformation of $AdS_4\times S^7$ can be recast into the same form as
(\ref{4d_general}) with the solution of three dimensional Toda equation given by (\ref{super4})
with an specific boundary condition at $y=0$ which is 
\bea
\theta=0&:&\;\;\;\;x>x_0,\;\;\;\;\;e^D\sim y\cr
{r}=0&:&\;\;\;\;x<x_0,\;\;\;\;\partial_yD\sim q/\sin^2(\theta),
\eea
where $x_0$ is a constant function of $Q$. This means that the solution is singular as 
expected.

\subsection{Penrose limit of superstar solution}

In this subsection we will derive the plane wave limit of the superstar solutions we have studied 
using the procedure we have developed in the previous section. Let us start with giant
graviton deformation of $AdS_7\times S^4$ case. To compare the result
with the $AdS_7\times S^4$ case we first rescale the coordinates as
$L_7=2/L,\;\;{\hat r}=Lr,\;Q=4qL^{-4}$ in which the solution is given by
\be
x=g(\hat{r})\cos\theta , \;\;\;\;\;  4y= L^{-3}\hat{r}^2\sin\theta , \;\;\;\;\;  
e^D = \frac{\hat{r}^2 f(\hat{r})}{4 L^6 g^2},
\ee
where $f=1+\hat{r}^2/4 + q/\hat{r}^2$.

To get the Penrose limit of the solution we will consider a limit in which
$L\rightarrow 0$ while keeping the following quantities fixed
\be
r_5= \frac{\hat{r}}{ L^3} , \;\;\;\;\; r_2 = \frac{\theta}{L^3}, \;\;\;\;\;\; {\tilde y}=\frac{y}{L^6}.
\ee
Of course one needs to scale $q$ properly. Depending on how to rescale $q$ we will get different
plane wave solutions. In particular if we rescale $q$ as $q={\tilde q}L^n$ for $n>6$, the limit will
remove the effect of the deformation such that the resulting solution is the standard plane
wave solution in 11-dimensions given by (\ref{mthpp}). On the other hand for $n<6$, the limit is not
well define. Finally for the case of $q={\tilde q}L^6$ in the limit of $L\rightarrow 0 $ we find
\bea
{\tilde x}_2&=&\frac{r_5^2}{4}-\frac{r_2^2}{2} -\frac{1}{4}{\tilde q}(\ln\frac{r_5^2+\tilde{q}}{4}-1),\cr
{\tilde y}&=&\frac{1}{4}r_5^2 r_2, \cr
e^D &=& \frac{\tilde{q}}{4} + \frac{r_5^2}{4},
\label{pp-singular}
\eea
where ${\tilde x}$ is fixed and defined by
\be
{\tilde x}_2=\frac{x-1 +\frac{3}{2} L^6 \ln L }{L^6}.
\ee
Note that unlike the $AdS_7$ case one needs to regularize ${\tilde x}$ as well. 

Let us now study the Penrose limit of the deformation of $AdS_4\times S^7$ solution. To do this 
we first rescale the coordinates as $L_4=1/L,\;\hat{r}=2rL,\;Q=2qL^{-1}$ in which the 
corresponding solution reads
\be
x=g(\hat{r})\cos\theta_1,\;\;\;\;\;\;y=\frac{1}{2L^3}\hat{r}\sin^2\theta,\;\;\;\;\;\;
e^D=\frac{4f(\hat{r})\sin^2\theta}{L^6 g^2(\hat{r})},
\ee
where $f=1+\hat{r}^4/4+q\hat{r}$.

To find the corresponding plane wave solution we will consider a limit in which $L\rightarrow 0$ while
keeping the following quantities fixed
\be
r_5=\frac{2\theta}{\sqrt{2}L^3},\;\;\;\;\;\;\;r_2=\frac{\sqrt{2}r}{4L^3},\;\;\;\;\;\;\;\tilde{y}=\frac{y}{L^6}
\tilde{x}_2=\frac{x-\sqrt{2}}{L^6}.
\ee
Moreover in order to get a finite solution one also needs to rescale $q$ appropriately For example one may 
rescale $q$ as $\tilde{q}=q/L^4$. Doing so we arrive at
\be
y=\frac{1}{\sqrt{2}}r_5^2r_2,\;\;\;\;\;\;\;\;\;\tilde{x}_2=\sqrt{2}(\frac{r_2^2}{2}-\frac{r_5^2}{4}),
\;\;\;\;\;\;\;\;e^D=r_5^2,
\ee
which leads to the same plane wave solution as that can be obtained from $AdS_4\times S^7$. We note that
as long as we require that the limit we are taking remains well defined, the results will be $q$-independent 
and it leads to the 11-dimensional non-singular plane wave solution. It is, of course, well-known,
that Penrose limit can remove singularity \cite{Alishahiha:2002ev}.

\section{Singular solutions and boundary condition}

In the previous sections we have studied several solutions, both singular and smooth,  of the three 
dimensional Toda equation. From what we have learned so far
we would like to study any possible boundary condition one 
might have for both singular and smooth solutions. 

The discussion of boundary condition for smooth 1/2 BPS solution in M-theory has recently 
been studied in \cite{Lin:2005nh} and the aim of this section is to further study the boundary
conditions. To do that following \cite{Lin:2004nb} we introduce a new set of coordinates
$\eta$ and $\rho$ such that
\be
e^D=\rho^2,\;\;\;\;\;\;\;y=\rho\partial_\rho V,\;\;\;\;\;\;\;x_2=\partial_\eta V,
\label{change}
\ee
for given function $V(\eta,\rho)$. Here we have assumed that the solution has 
translation Killing vector in $x_1$ direction. One may also consider the case 
where we have rotation Killing vector in $\phi$ direction in the $(r,\phi)$ plane. In this case
the coordinate $x_2$ in  (\ref{change}) can be defined by $r=e^{x_2}$ in which we will also need
to shift $D$ by $2\ln x_2$.

In this new coordinates system the three dimensional Toda equation reduces to a Laplace equation
with the cylindrically symmetry for function $V$
\be
\frac{1}{\rho}\partial_\rho(\rho V)+\partial_\eta^2V=0.
\ee
One then needs to understand the corresponding boundary conditions (\ref{boundary1}) in terms of these new
coordinates which must be translated to boundary conditions for $V$ as a function of $\rho$ and $\eta$.

Actually it is shown \cite{Lin:2005nh} that finding a solution for the Toda equation reduces 
to an auxiliary three dimensional electrostatic system with conducting disks setting at 
positions $\eta_i$ with radii $\rho_i$. Let us first review the construction of the auxiliary system.

Starting from the definition of $\rho$ in terms of $D$ one finds
\be
\rho^2\partial_yD=\frac{-2\partial_\eta^2V}{(\partial_{\eta\rho}^2V)^2+(\partial_\eta^2V)^2},
\ee
from which we can show that $\partial_\rho V\rightarrow 0$ whenever $\rho \rightarrow 0$. This can be
seen as follows. For $\rho\rightarrow 0$ one finds $\partial_\eta^2V\rightarrow 0$ which can be used
to conclude $\frac{1}{\rho}\partial_\rho(\rho\partial_\rho V)\rightarrow 0$ which leads to
$\partial_\rho V\sim\rho\rightarrow 0$. As the conclude we note that imposing boundary condition
at $y\rightarrow 0$ can be expressed in the new coordinates system as the limit where $\rho\partial_\rho V
\rightarrow 0$ which is possible if
\be
\rho\neq 0,\;\;\partial_\rho V\rightarrow 0,\;\;\;\;\;\;\;{\rm or}\;\;\;\;\;\;\rho\rightarrow 0,\;
\partial_\rho V\rightarrow 0
\ee
One can see that the first condition is consistent with the boundary condition in which
$\partial_yD=0$ while the second one is consistent with the boundary condition given by $e^D\sim y$ as
$y\rightarrow 0$. In the first case we also get $\partial_\rho^2V=0$ and therefore the curve $y=0$,
$\rho\neq 0$ is at constant values of $\eta$. Moreover if we interpret $V$ as an electrostatic potential,
then $-\partial_\rho V=0$ is the electric field along $\rho$ and the condition that it is zero corresponds
to the presence of a charged conducting surface. In general we could have $n$ conducting disks with
radii $\rho_i$ sitting at $\eta_i$ for $i=1,\cdots ,n$ which could be in an external electric field.
The second condition just tell us that the potential $V$ is regular at $\rho=0$.

The superstar solutions we have studied in the previous section could give us an insight how to change 
the boundary condition if we want to relax the smoothness of the solution. 
One way to do this is just to remove the regularity condition
at $\rho=0$ which from three dimensional auxiliary electrostatic system it means we should have a charge
distribution at $\rho=0$.  Let us now consider the superstar case studied in previous section in more detail.

Consider a boundary condition in which at $y\rightarrow 0$ we have $e^D\sim a+by$ for some non zero
$y$-independent parameters $a$ and $b$. In the $(\rho,\eta)$ coordinates it means that we have 
the case where $\rho\neq 0$ while $\partial_\rho V\rightarrow 0$ for $y\rightarrow 0$. On the other hand 
for the other boundary condition we may still insist to get $\partial_yD=0$ at $y=0$ which again
can be satisfied if we impose $\rho\neq 0$ while $\partial_\rho V\rightarrow 0$ boundary condition.
As a conclusion we note that this boundary condition is enough to get whole supergravity solution though
the solution would be singular. From there dimensional electrostatic system point of view, this corresponds
to the case where the potential $V$ is not regular at $\rho=0$ which means that we should have a charge
distribution at $\rho=0$. 

To see how this could happen it is instructive to study the singular plane wave solution we have 
obtained in the previous section (\ref{pp-singular}) which  can be recast to the following form
\be
e^D=\rho^2,\;\;\;\;\;\;\;\;y=2\eta(\rho^2-\frac{\tilde{q}}{4}),\;\;\;\;\;\;x_2=(\rho^2-\frac{\tilde{q}}{2}\ln\rho)-
2\eta^2.
\ee
Note that to get the above expression we used a change of coordinates as $r_5^2+\tilde{q}=4\rho^2,\;r_2=2\eta$.
First of all we observe that in the limit of $y\rightarrow 0$ we get boundary conditions for $D$ which which
are 
\be   
e^D\sim \frac{\tilde{q}}{4},\;\;\;{\rm for}\;\;\;\rho^2\rightarrow \frac{\tilde{q}}{4},\;\;\;\;\;\;\;\;\;\;
\partial_yD=0 \;\;\;{\rm at }\;\;\;\eta=0.
\ee
It is easy to see that this solution can be
obtained from the following potential
\bea
\label{poten1}
V=\rho^2\eta -\frac{2}{3}\eta^3-\frac{\tilde{q}}{2}\eta \ln\rho.
\eea
We see that in $\rho\rightarrow 0$ the potential blows up as $\ln \rho$ which can be interpreted as
a potential produced by a distribution of a linear charge setting at $\rho=0$ and $\tilde{q}\eta$
is the total electric charge. 

One may also modify the other boundary condition to get a possible singular solution. More precisely
we consider a boundary condition in which at $y=0,\;\; \partial_yD\neq 0$, while the other condition
remains unchanged, namely  for $y\rightarrow 0$ one has $e^D\sim y$.
To understand this boundary condition and the corresponding singular solution better let us
study an explicit example exhibits such a singular boundary condition. 

Consider a solution given by a potential corresponding to a charged ring located at the origin. 
Its electrostatic potential is given by
\be
V_0=\frac{a}{\sqrt{\eta^2+\rho^2}}.
\ee
We may consider a background potential which might present because of an infinite or finite disk
at $\eta=0$. For example in the case with an infinite disk the whole system is given by the
following potential
\be
V=\rho^2\eta -\frac{2}{3}\eta^3+\frac{a}{\sqrt{\eta^2+\rho^2}},
\ee
and therefore we arrive at
\be
e^D=\rho^2,\;\;\;\;\;\;\;y=2\rho^2\eta-\frac{a\rho^2}{(\eta^2+\rho^2)^{3/2}},\;\;\;\;\;\;
x=\rho^2-2\eta^2-\frac{a\eta}{(\eta^2+\rho^2)^{3/2}}.
\ee
This leads to a singular solution such that $e^D\sim y$ for $y\rightarrow 0$ while $\partial_yD\neq 0$ at
$y=0$ which is the singular boundary condition we were interested in. Therefore the solution
is singular at $y=0$. This solution has also been studied in \cite{Bak:2005ef}.

Starting from this potential one may generate other solutions which could be either singular or smooth.
In fact one can easily show that the following potential 
\be
V=\rho^2\eta -\frac{2}{3}\eta^3+V_n,
\label{potential}
\ee
with 
\be 
V_n=\frac{\partial^n}{\partial\eta^n}\left(\frac{a}{\sqrt{\eta^2+\rho^2}}\right)
\ee
satisfies the three dimensional Laplace equation. Actually this is multipoles expansion for an 
electrostatic potential. Using this one finds
\be
e^D=\rho^2,\;\;\;\;\;y=\rho^2(2\eta-\partial_\eta^n\frac{a}{(\eta^2+\rho^2)^{3/2}}),
\;\;\;\;x_2=\rho^2-2\eta^2+\partial_\eta^{n+1}\frac{a}{(\eta^2+\rho^2)^{1/2}},
\label{so}
\ee 
from which at $y\rightarrow 0$ one has $e^D\sim y$. On the other hand we have
\be
\partial_\eta^2V=-4\eta+\partial_\eta^{n+2}\frac{a}{(\eta^2+\rho^2)^{1/2}},
\ee
which together with (\ref{so}) shows that for odd $n$ the $y=0$ has a solution at $\eta=0$ which also
solves $\partial_\eta^2V=0$ and therefore one get $\partial_yD=0$ at $y=0$. Thus the potential
(\ref{potential}) leads to smooth solutions for odd $n$. In particular for $n=1$ is the leading
asymptotic form of the solution for the plane wave matrix model \cite{Lin:2005nh}. 
In general we can find a smooth solution by summing up $2^{2k+1}$-multipoles
\be
V=\rho^2\eta -\frac{2}{3}\eta^3+\sum_{k=0}\partial^{2k+1}_\eta\frac{a}{(\eta^2+\rho^2)^{1/2}},
\ee 
while for a potential coming from a summation over $2^{2k}$-multipoles we get a singular 
solution such that at $y=0,\;\partial_yD\neq 0$. 

Another set of singular solutions have also been studied in \cite{Bak:2005ef}
where the solution has null singularity. These solutions can simply be obtained by considering
an ansatz such that $V=\phi(\rho)T(\eta)$ which leads to the following solutions for the
three dimensional Laplace equation
\be
K_0(k\rho)e^{ik\eta},\;\;\;\;I_0(k\rho)e^{ik\eta},\;\;\;\;J_0(k\rho)e^{k\eta},\;\;\;\;
N_0(k\rho)e^{k\eta}.
\ee
All of these solutions which are given in terms of Bessel function are singular. Nevertheless
one may construct new smooth solutions using the same procedure. For example consider the following
solution of the three dimensional Laplace equation
\be
V=\rho^2\eta-\frac{2}{3}\eta^3+I_0(k\rho)\sin(k\eta),
\ee
which leads to
\be
e^D=\rho^2,\;\;\;\;\;y=\rho\bigg{(}2\rho\eta-kI_1(k\rho)\sin(k\eta)\bigg{)},\;\;\;\;\;x_2=\rho^2-2\eta^2+kI_0(k\rho)\cos(k\eta).
\ee
The boundary condition of this solution is as follows. First of all we note that at $y=0$ we have two
solutions either $\rho=0$ or $\eta=0$ and in fact we find
\be
e^D\sim y\;\;\;\;\;{\rm for}\;\;\rho\rightarrow 0,\;\;\;\;\;\;\;\;\;\partial_yD=0\;\;\;\;\;{\rm at}\;\; \eta=0.
\ee
This solution (without the plane wave part) and its compactification to type IIA string theory 
has been studied in \cite{Lin:2005nh} where in type IIA string theory this corresponds to
wrapped NS5-brane on $S^5$. One could also consider a solution given by $J_0(k\rho)\sinh(k\eta)$ 
which has the same behavior a above. 
If we had considered $\cos(k\eta)$ or $\cosh(k\eta)$  we would have gotten solutions which were singular. In fact
in these cases we get $e^{D}\sim y$ when $y\rightarrow 0$ while at $y=0$ one finds $\partial_yD\neq 0$.
On the other hand if we consider $N_0(k\rho)\sinh(k\eta)$ or $K_0(k\rho)\sin(k\eta)$ then solutions 
will also be singular with boundary condition given by  $e^D\sim a+by$ for $y\rightarrow 0$ while the other 
boundary condition remains the same as before. In the next section we will study compactifcation of solutions
given by these potential following \cite{Lin:2005nh}. 

Finally we note that starting from two smooth solutions it is possible to find a singular
solution one. For example consider a solution given by potential $V=I_0(k\rho)\cos(k\eta)$.
It is non-singular and in fact is equivalent to the case studied in\cite{Lin:2005nh}.  But considering
this potential to the background potential representing an infinite conducting disk at 
$\eta=0$ leads  to a singular solution such that for $y\rightarrow 0$ goes as $e^{D}\sim
{\rm constant}$.

\section{Ten dimensional solutions}

In the previous section we have studied both singular and smooth 1/2 BPS solutions of M-theory.
All solutions we have considered have an isometry which is translation invariant along $x_1$  and 
therefore can be compactified to find type IIA solutions. For a general background, doing so 
one finds \cite{Lin:2005nh}
\bea
ds^2&=&\bigg{(}\frac{{\ddot V}-2{\dot V}}{-V''}\bigg{)}^{1/2}\bigg{[}
\frac{-4{\ddot V}}{{\ddot V}-2{\dot V}}dt^2+\frac{-2V''}{{\dot V}}(d\rho^2+d\eta^2)+4d\Omega_5^2+
\frac{2V''{\dot V}}{\Delta}d\Omega_2^2\bigg{]},\cr &&\cr
e^{4\phi}&=&\frac{4({\ddot V}-2{\dot V})^2}{-V''{\dot V}^2\Delta^2},\;\;\;\;\;\;\;\;C_3=
-4\frac{{\dot V}^2V''}{\Delta}dt\wedge d^2\Omega,\cr &&\cr
C_1&=&-\frac{2{\dot V}'{\dot V}}{{\ddot V}-2{\dot V}}dt,\;\;\;\;\;\;\;\;
B_2=2\left(\frac{{\dot V}'{\dot V}}{\Delta}+\eta
\right)d^2\Omega,
\eea
where $\Delta=({\ddot V}-2{\dot V})V''-{\dot V}'^2$ and the dots denote derivatives with respect to
$\log\rho$ and the primes denote derivatives with respect to $\eta$.

In principle one can start from the potentials we have written in the previous
section and plugging them into
the above expression to get type IIA supergravity solutions. For example plugging $V=I_0(k\rho)\sin(k\eta)$
we will get type IIA NS5-brane wrapped on $S^5$ \cite{Lin:2005nh}. Different aspects of this
solution using semi-classical closed strings has been studied in \cite{Ebrahim:2005sb}.

As another example let us consider the solution corresponding to singular plane wave solution,
given by the potential $V=\rho^2\eta-\frac{2}{3}\eta^3-2q^2\eta\ln\rho$. The solution is 
\bea
ds^2 &=& -\frac{4}{q}\rho^2 dt^2 + \frac{4q}{\rho^2-q^2} (d\rho^2+d\eta^2)+ 4qd\Omega_5^2+ \frac{4q\eta^2(\rho^2-q^2)}{4q^2\eta^2+(\rho^2-q^2)} d\Omega_2^2, \cr\cr
e^{4\phi} &=& \frac{q^6}{(\rho^2-q^2)(4q^2\eta^2+(\rho^2-q^2)^2)^2},\;\;\;\;C_3 = -\frac{16\eta^3(\rho^2-q^2)}{4q^2\eta^2+(\rho^2-q^2)^2} dt \wedge d^2\Omega, 
 \cr\cr
 B &=&2(-\frac{(\rho^2-q^2)\eta}{4q^2\eta^2+(\rho^2-q^2)^2}+\eta)d^2\Omega,\;\;\;
 C_1 =-\frac{2}{q^2}(\rho^2-q^2)^2 dt.
\label{SS}
\eea 
This solution and the corresponding boundary condition are very similar to that 
studied in \cite{Lin:2005nh} in case of ${\cal N}=4$ SY theory on $R\times S^3/Z_k$. In that
case the potential was given by $V=\rho^2-2\eta^2-2q^2\ln\rho$ leading to the following solution
\bea
ds^2 &=& -\frac{4}{q}\rho^2 dt^2 + \frac{4q}{\rho^2-q^2} (d\rho^2+d\eta^2)+ 4qd\Omega_5^2+ 
\frac{\rho^2-q^2}{q} d\Omega_2^2, \cr\cr
e^{2\phi} &=& \frac{q}{4(\rho^2-q^2)},\;\;\;\;
C_3 = -\frac{4(\rho^2-q^2)}{q^2} dt \wedge d^2\Omega,\;\;\;
 B =2\eta d^2\Omega.
 \eea 
It would be interesting to find the gauge theory dual of the solution (\ref{SS}).

One can also consider other potential we have presented in previous section to find type IIA solutions.
We note, however, given a potential of the three dimensional auxiliary electrostatic system is not 
enough to get a 10-dimensional physical solution. One needs to check whether the different components
of the metric are positive-definite which is given by the following conditions \cite{Lin:2005nh}
\be
\Delta \leq 0,\;\;\;\;\;V''\leq 0,\;\;\;\;\; {\ddot V}-2{\dot V}\geq 0, \;\;\;\;\;{\dot V}\geq 0.
\ee
For example if we start from a potential given by $V=I_0(k\rho)\sin(k\eta)$ one can see that it 
satisfies the above conditions and in fact lead to a 10-dimensional solution representing type IIA 
NS5-brane wrapping on $S^5$. On the other hand if we start from other solutions in this kind,
such that $K_0(k\rho)\sin(k\eta)$, it does not satisfies the above conditions and therefore
does not lead to a well-defined 10-dimensional solution. Nevertheless it is possible to add 
a background potential to make the solution well-defined. For example we can consider the 
background potential given by $V_b=\rho^2\eta-\frac{2}{3}\eta^3$.

Since the superstar solutions of M-theory have also an isometry along angular direction
$\psi$ it would also be interesting to compactify these solutions along this direction
to find their type IIA string theory counterpart. Actually starting from superstar solution in 
$AdS_7 \times S^4$ (\ref{metric0}) one arrives at
\bea
ds^2 &=& 2H_1^{1/2}\cos\theta_1  \bigg[-4 \frac{f}{H_1}dt^2 + 4f^{-1} dr^2  + 4r^2 d\Omega_5^2 
+ L_7^2 \left(d\theta_1^2 +
 \frac{\sin^2\theta_1}{\Delta} d\Omega_2^2\right)\bigg], \cr
e^{2\phi} &=&g_s^2 \frac{H_1^{3/2} \cos^3\theta_1}{8\Delta} ,\;\;\;\;\;C_1 = -\frac{1}{g_s} H_1^{-1} dt,
\eea
while for  $AdS_4\times S^4$ we get
\bea
ds^2 &=& 2 \Delta^{1/2}H_1^{1/2}\cos\theta_1 \bigg[- \frac{f}{H_1}dt^2 + f^{-1}
 dr^2 +  r^2 d\Omega_2^2 + 4  L_4^2\left( d\theta_1^2 
+ \frac{\sin^2\theta _1}{\Delta}d\Omega_5^2\right) \bigg], \cr
e^{2\phi} &=& g_s^2\frac{8 H_1^{3/2} \cos^3\theta_1}{\Delta^{1/2}} \;\;\;\;\;\;
C_1 = -\frac{1}{g_s} H_1^{-1} dt.
\eea
In addition we have also a RR 3-form as well as non-zero B-field.

These can be thought of as type IIA superstar solutions. The first one might be interpreted as D2-brane
solution wrapped on $S^2$ deformed by dual giant graviton  given by spherical D2-branes 
wrapping the $S^2$ sphere. On the other hand the second one corresponds to NS5-brane wrapped on $S^5$ 
deformed by dual giant graviton given by spherical NS5-brane wrapping the $S^5$ sphere. 

\section{Conclusions}

In this paper we have studied different aspects of 1/2 BPS geometries in M-theory using
some known M-theory supergravity solutions. These include both singular and smooth solutions.
All of these solutions can fully be given by a function $D(y,x_1,x_2)$ which satisfies a three 
dimensional Toda equation with specific boundary condition at $y=0$. Depending on how
the function $D$ behaves at $y=0$ one can assign a droplet in $(x_1,x_2)$-plane for any
solution. 

Since the solution is invariant under a conformal map acting as $x_1+ix_2\rightarrow f(x_1+ix_2)$ and
$D\rightarrow D-\ln |\partial f|^2$ the shape of the droplet does not play an essential role, though
the topology of the plane is important. In particular we have seen for both $AdS_{4,7}\times S^{7,4}$ and
their corresponding plane wave solution we get the same shape in the $(x_1,x_2)$-plane 
using a conformal map. Of course we note that the topology of the plane for these cases
are different and from that one may specify the correct solution.

We have also seen how one can define a boundary condition to get singular but physically acceptable 
solutions by making use of known M-theory superstar solutions. We have also studied the penrose limit
of these singular solutions using LLM parameterization.

It is also shown that for solutions with translation invariant one may map the Toda equation to a
three dimensional Laplace equation with a function $V$ which could be interpreted as a potential
of electrostatic system in three dimensions. Using this  we have reformulated the singular
solutions in this language and how to generalized to get new solutions. Since all of these 
solutions have an isometry one could also compactify them to get type IIA solutions which might
provide the gravity dual for some gauge theories. It would be interesting if one could find and 
study these guage theories.

\noindent\textbf{Acknowledgments}

We would like to thank A.E. Mosaffa and M.M. Sheikh-Jabbari for useful discussions.


\begin{thebibliography}{99}


\bibitem{Maldacena:1997re}
  J.~M.~Maldacena,
   ``The large $N$ limit of superconformal field theories and supergravity,''
 Adv.\ Theor.\ Math.\ Phys.\  {\bf 2}, 231 (1998)
  [Int.\ J.\ Theor.\ Phys.\  {\bf 38}, 1113 (1999)]
  [arXiv:hep-th/9711200].


\bibitem{Gubser:1998bc}
  S.~S.~Gubser, I.~R.~Klebanov and A.~M.~Polyakov,
   ``Gauge theory correlators from non-critical string theory,''
    Phys.\ Lett.\ B {\bf 428}, 105 (1998)
  [arXiv:hep-th/9802109].

\bibitem{Witten:1998qj}
  E.~Witten,
   ``Anti-de Sitter space and holography,''
  Adv.\ Theor.\ Math.\ Phys.\  {\bf 2}, 253 (1998)
  [arXiv:hep-th/9802150].



\bibitem{BMN}
  D.~Berenstein, J.~M.~Maldacena and H.~Nastase,
  ``Strings in flat space and pp waves from ${\cal N} = 4$ super Yang Mills,''
  JHEP {\bf 0204}, 013 (2002)
  [arXiv:hep-th/0202021].

\bibitem{Gubser:2002tv}
S.~S.~Gubser, I.~R.~Klebanov and A.~M.~Polyakov, ``A
semi-classical limit of the gauge/string correspondence,'' Nucl.\
Phys.\ B {\bf 636}, 99 (2002) [arXiv:hep-th/0204051].

\bibitem{Frolov:2002av}
S.~Frolov and A.~A.~Tseytlin, ``Semiclassical quantization of
rotating superstring in $AdS_5\times S^5$,'' JHEP {\bf 0206}, 007
(2002) [arXiv:hep-th/0204226].



\bibitem{Lin:2004nb}
  H.~Lin, O.~Lunin and J.~Maldacena, ``Bubbling AdS space and 1/2 BPS geometries,''
  JHEP {\bf 0410}, 025 (2004)
  [arXiv:hep-th/0409174].
  
  

\bibitem{Suryanarayana:2004ig}
  N.~V.~Suryanarayana,
  ``Half-BPS giants, free fermions and microstates of superstars,''
  arXiv:hep-th/0411145.



\bibitem{Bak:2005ef}
  D.~Bak, S.~Siwach and H.~U.~Yee,
  ``1/2 BPS geometries of M2 giant gravitons,''
  Phys.\ Rev.\ D {\bf 72}, 086010 (2005)
  [arXiv:hep-th/0504098].



\bibitem{MacConamhna:2005vt}
  O.~A.~P.~Mac Conamhna,
  ``The geometry of extended null supersymmetry in M-theory,''
  arXiv:hep-th/0505230.




\bibitem{Spalinski:2005ha}
  M.~Spalinski,
  ``Some half-BPS solutions of M-theory,''
  arXiv:hep-th/0506247.


\bibitem{Lin:2005nh}
  H.~Lin and J.~Maldacena,
  ``Fivebranes from gauge theory,''
  arXiv:hep-th/0509235.

\bibitem{Milanesi:2005tp}
  G.~Milanesi and M.~O'Loughlin,
  JHEP {\bf 0509}, 008 (2005)
  [arXiv:hep-th/0507056].

\bibitem{Ganjali:2005cr}
  M.~A.~Ganjali,
  ``On Toda equation and half BPS supergravity solution in M-theory,''
  arXiv:hep-th/0511145.



\bibitem{Alishahiha:2005fc}
  M.~Alishahiha, H.~Ebrahim, B.~Safarzadeh and M.~M.~Sheikh-Jabbari,
  ``Semi-classical probe strings on giant gravitons backgrounds,''
  JHEP {\bf 0511}, 005 (2005)
  [arXiv:hep-th/0509160].



\bibitem{Berenstein:2002jq}
  D.~Berenstein, J.~M.~Maldacena and H.~Nastase,
  ``Strings in flat space and pp waves from ${\cal N} = 4$ super Yang Mills,''
  JHEP {\bf 0204}, 013 (2002)
  [arXiv:hep-th/0202021].

\bibitem{Blau:2002dy}
  M.~Blau, J.~Figueroa-O'Farrill, C.~Hull and G.~Papadopoulos,
  ``Penrose limits and maximal supersymmetry,''
  Class.\ Quant.\ Grav.\  {\bf 19}, L87 (2002)
  [arXiv:hep-th/0201081].


\bibitem{Takayama:2005bc}
  Y.~Takayama and K.~Yoshida,
  ``Bubbling 1/2 BPS geometries and Penrose limits,''
  Phys.\ Rev.\ D {\bf 72}, 066014 (2005)
  [arXiv:hep-th/0503057].


\bibitem{Ebrahim:2005uz}
  H.~Ebrahim and A.~E.~Mosaffa,
  ``Semiclassical string solutions on 1/2 BPS geometries,''
  JHEP {\bf 0501}, 050 (2005)
  [arXiv:hep-th/0501072].


\bibitem{Leblond:2001gn}
  F.~Leblond, R.~C.~Myers and D.~C.~Page,
  ``Superstars and giant gravitons in M-theory,''
  JHEP {\bf 0201}, 026 (2002)
  [arXiv:hep-th/0111178].



\bibitem{Balasubramanian:2001dx}
  V.~Balasubramanian and A.~Naqvi,
  ``Giant gravitons and a correspondence principle,''
  Phys.\ Lett.\ B {\bf 528}, 111 (2002)
  [arXiv:hep-th/0111163].

\bibitem{Cvetic:1999xp}
  M.~Cvetic {\it et al.},
   ``Embedding AdS black holes in ten and eleven dimensions,''
  Nucl.\ Phys.\ B {\bf 558}, 96 (1999)
  [arXiv:hep-th/9903214].


\bibitem{Alishahiha:2002ev}
  M.~Alishahiha and M.~M.~Sheikh-Jabbari,
  ``The pp-wave limits of orbifolded $AdS_5\times S^5$,''
  Phys.\ Lett.\ B {\bf 535}, 328 (2002)
  [arXiv:hep-th/0203018].


\bibitem{Ebrahim:2005sb}
  H.~Ebrahim,
  ``Semiclassical strings probing NS5 brane wrapped on $S^5$,''
  arXiv:hep-th/0511228.
\end{thebibliography}
\end{document}